\documentclass[aps,twocolumn,preprintnumbers]{revtex4}

\usepackage{graphicx}  
\usepackage{subfigure}
\usepackage{multirow}
\usepackage{hyperref}
\usepackage[dvipsnames]{xcolor}
\usepackage{changepage}
\usepackage{longtable}
\usepackage{parskip}
\usepackage[T1]{fontenc}
\usepackage{dcolumn}   
\usepackage{amsfonts}  
\usepackage{amsmath}   
\usepackage{amssymb}   
\usepackage[twoside]{fancyhdr}

\newcommand{\pwisein}{\left\{ \begin{array}{ll}}
\newcommand{\pwiseout}{\end{array}\right.}

\begin{document}

\title{Assessing the role of threshold conditions in the determination of uncertainties in pole extractions using Padé approximants
}

\author{Balma Duch and Pere Masjuan}

\affiliation{Grup de Física Teòrica, Departament de Física, 
Universitat Autònoma de Barcelona, 08193 Bellaterra (Barcelona), Spain}

\affiliation{Institut de Física d’Altes Energies (IFAE) and 
The Barcelona Institute of Science and Technology (BIST), 
Campus UAB, 08193 Bellaterra (Barcelona), Spain}

\begin{abstract} \noindent In this letter, we discuss the determination of the $f_0(500)$ resonance by analytic continuation through Padé approximants of the $\pi\pi$-scattering amplitude from the physical region to the pole in the complex energy plane. Using as input a class of admissible parametrizations of the scalar-isoscalar $\pi\pi$ partial wave and imposing now the correct threshold behavior of the partial amplitude, we improve on the determinations of pole positions obtained in Ref. \cite{Caprini:2016uxy}, thus empowering the \textit{Padé method} as a simple and precise tool for extracting resonance poles from amplitudes.
\end{abstract} 

\maketitle 
\section{Introduction}
\noindent
The determination of a broad resonance such as the $I = J = 0$ lowest state $f_0(500)$ has long been a notoriously difficult problem. Its associated S-matrix pole lies deep in the complex energy plane, yielding the analytic extraction of its position a non-trivial task. For many years, the limited precision of low-energy $\pi\pi$ scattering data led to large uncertainties in the extracted resonance parameters, to the point that the very existence of the state was debated \cite{Pelaez:2015qba}. A major breakthrough was achieved through the use of dispersion relations \cite{Colangelo:2001df, Ananthanarayan:2000ht}, in particular Roy \cite{Roy:1971tc} and GKPY equations \cite{Garcia-Martin:2011iqs}, which provide a stable framework for analytic continuation and have led to precise determinations of the $\sigma$ pole. Dispersive approaches have been proved to be a very successful tool to obtain precise determinations of phase shifts and pole parameters in several instances \cite{Colangelo:2001df,Buettiker:2003pp,Caprini:2005zr,Garcia-Martin:2011nna}. However, they are based on a complicated although powerful machinery which makes them difficult to use
except for a limited number of cases.

Alternative analytic continuation techniques have been developed to extract resonance properties, extending their application from the light meson sector to baryon spectroscopy. These include the Pietarinen expansion method \cite{Svarc:2013laa}, chiral unitary approaches based on the N/D method \cite{Meissner:1999vr}, and continued fractions \cite{Pelaez:2025jrn}. Notably, these techniques share the same analytical principles as Padé approximants (PAs). Continued fractions are a particular way to represent diagonal PAs \cite{BakerGraves-Morris1996}, and also N/D methods can be understood from the point of view of PAs \cite{Masjuan:2008cp}. Furthermore, recent non-parametric approaches have also been proposed \cite{Salg:2025now}. In the present work, we focus on PAs as a robust model-independent tool to perform the analytic continuation of scattering amplitudes and extract resonance poles. As a novelty, we center our effort on the so-called 2 point-Padé approximants \cite{BakerGraves-Morris1996,Masjuan:2007ay,Masjuan:2008fr} so far never used in the context of the extraction of resonance pole.

To this end, we consider the $I = J = 0$ partial wave $t_{00}(s)$. By imposing the correct threshold behavior, we explore the complex plane of the amplitude to locate the pole position of the $f_0(500)$ meson. To do so, we build on previous works \cite{Caprini:2016uxy,Masjuan:2013jha,Masjuan:2014psa} where Padé Theory was used as an analytic tool to extend the amplitude from the real axis to the complex plane. In particular, PAs and related sets of convergence theorems were employed. PAs provide a powerful, model-independent method to extract resonance parameters, such as masses, widths, and residues, from the analytic continuation of elastic scattering amplitudes \cite{BakerGraves-Morris1996, Masjuan:2013jha}. These approximants are rational functions, $P^M_N(s,s_0)$, defined as the ratio of a polynomial of degree $N$ to another of degree $M$. They are constructed so that their Taylor expansion around a reference point $s_0$ exactly coincides with that of the function $F(s)$ up to the highest possible order, $\mathcal{O}((s-s_0)^{N+M})$. When derivatives of $F(s)$ at $s_0$ are known with sufficient precision, the PAs coefficients $a_k$ can be determined directly; alternatively, they may be extracted from experimental data by fitting over a suitable interval. A key advantage of PAs is their well-understood convergence properties. In the limit $N,M \to \infty$, the approximants converge to the target function under very general conditions within its range of convergence \cite{BakerGraves-Morris1996}, which ensures reliable resonance extraction and allows for a systematic estimation of theoretical uncertainties associated with truncating the PA sequence.  

The Montessus de Ballore's theorem \cite{Montessus1902} is particularly useful for resonance analysis. If $F(s)$ is analytic in a disk $B_\delta(s_0)$ except for a single pole at $s=s_p$, then the sequence of one-pole PAs
\begin{equation}
P_1^N(s,s_0) = \frac{\sum_{k=0}^{N-1} a_k (s-s_0)^k + a_N (s-s_0)^N}{1 - \frac{a_{N+1}}{a_N} (s-s_0)}
\end{equation}
converges to $F(s)$ for any compact subset excluding $s_p$, and its pole
\begin{equation}
s_p^{PA} = s_0 + \frac{a_N}{a_{N+1}}
\label{sP1N}
\end{equation}
approaches $s_p$ as $N \to \infty$. This allows one to unfold the second Riemann sheet and locate resonance poles in the complex plane in a model-independent way. If $F(s)$ contains $N^*$ poles within the disk, one may construct $P^M_N(s,s_0)$ with $N\ge N^*$. Extra poles in the approximant simulate more distant singularities, such as branch points generated by unitarity \cite{Masjuan:2008cp}.

The novelty presented in this letter is the use of 2-point PAs as an improvement to 1-point PAs used in Refs.\cite{Caprini:2016uxy,Masjuan:2013jha,Masjuan:2014psa}, to extract resonance poles. Our new PAs are then built combining the expansion at $s_0$ and the one at $\pi\pi$ production threshold. 
This second point heavily constrains the mobility of the analytic continuation driven by the higher derivatives of the parameterization and thus reduces the theoretical uncertainty on the pole determination from the PA method.

\begin{figure*}[]

  \begin{minipage}[b]{0.19\textwidth}
    \centering
    \includegraphics[width=\linewidth]{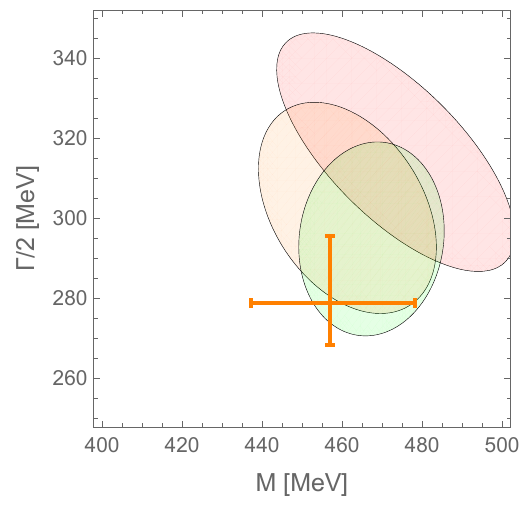}\\
    \small (a) $v_{1}$ 
  \end{minipage}\hfill
  \begin{minipage}[b]{0.19\textwidth}
    \centering
    \includegraphics[width=\linewidth]{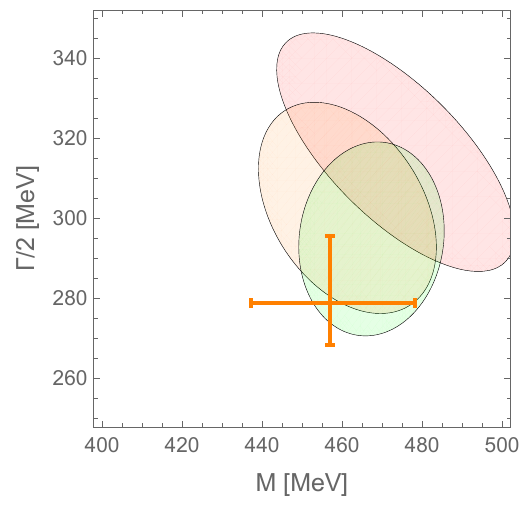}\\
    \small (b) $v_{2}$ 
  \end{minipage}\hfill
  \begin{minipage}[b]{0.19\textwidth}
    \centering
    \includegraphics[width=\linewidth]{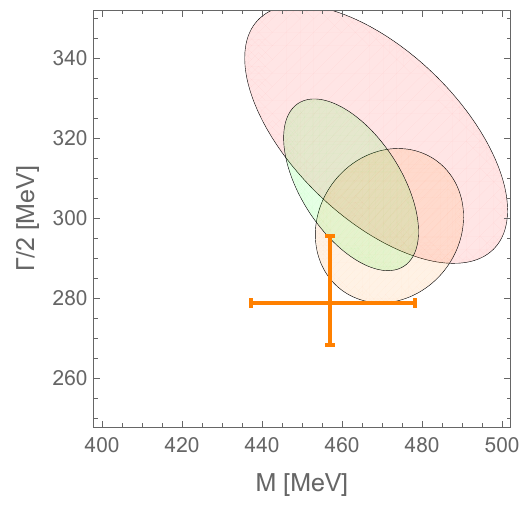}\\
    \small (c) $v_{3}$ 
  \end{minipage}\hfill
  \begin{minipage}[b]{0.19\textwidth}
    \centering
    \includegraphics[width=\linewidth]{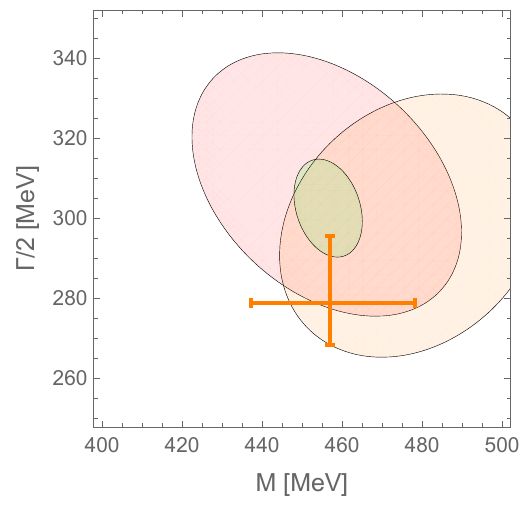}\\
    \small (d) $v_{4}$ 
  \end{minipage}\hfill
  \begin{minipage}[b]{0.19\textwidth}
    \centering
    \includegraphics[width=\linewidth]{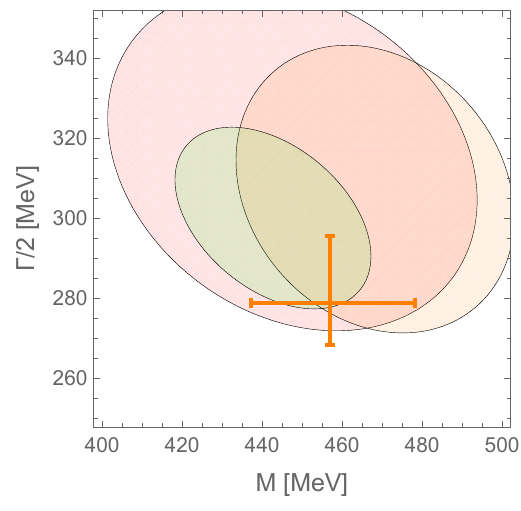}\\
    \small (e) $v_{5}$ 
  \end{minipage}

  \caption{Overlap of the 68\% CL ellipses for the $P_{1}^{N}$'s sequence pole position: red for $P_{1}^{2}$, orange for $P_{1}^{3}$, and green for $P_{1}^{4}$. Panels a) to e) correspond to parameterizations $v_1$-$v_5$. The orange cross is the reference number $s_p=(457^{+14}_{-13} - i 279^{+11}_{-7})$.}
  \label{P1iParametrization}
\end{figure*} 
\begin{figure*}[t!]

  \begin{minipage}[b]{0.19\textwidth}
    \centering
    \includegraphics[width=\linewidth]{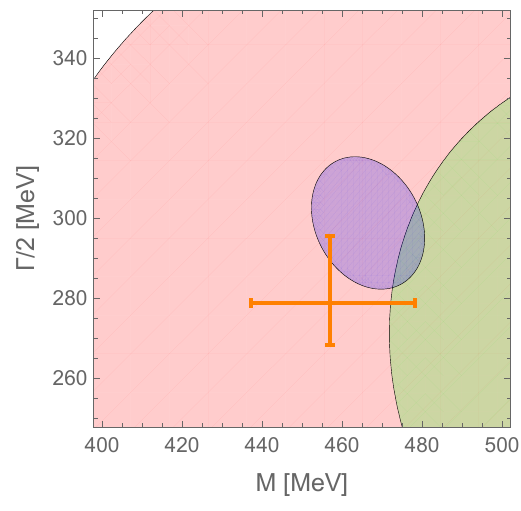}\\
    \small (a) $v_{1}$ 
  \end{minipage}\hfill
  \begin{minipage}[b]{0.19\textwidth}
    \centering
    \includegraphics[width=\linewidth]{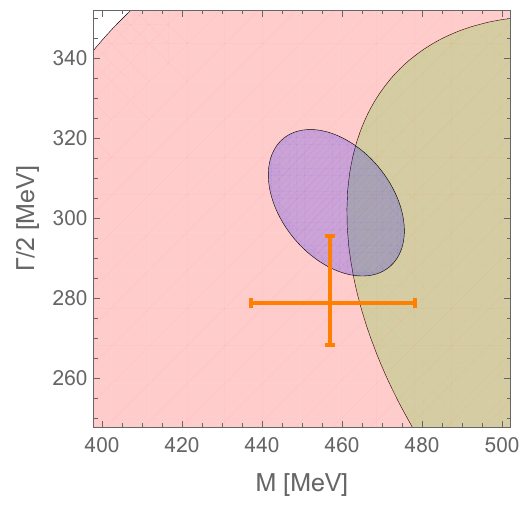}\\
    \small (b) $v_{2}$ 
  \end{minipage}\hfill
  \begin{minipage}[b]{0.19\textwidth}
    \centering
    \includegraphics[width=\linewidth]{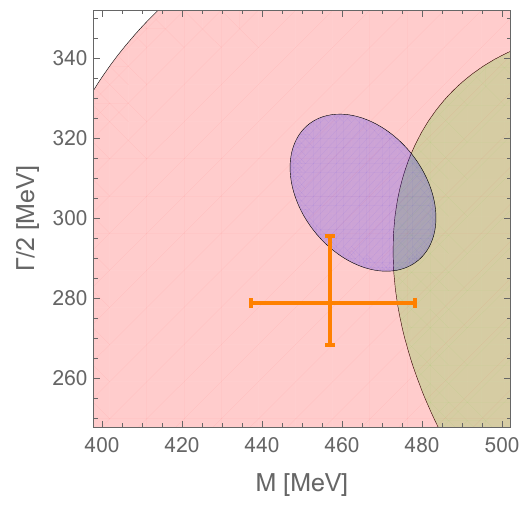}\\
    \small (c) $v_{3}$
  \end{minipage}\hfill
  \begin{minipage}[b]{0.19\textwidth}
    \centering
    \includegraphics[width=\linewidth]{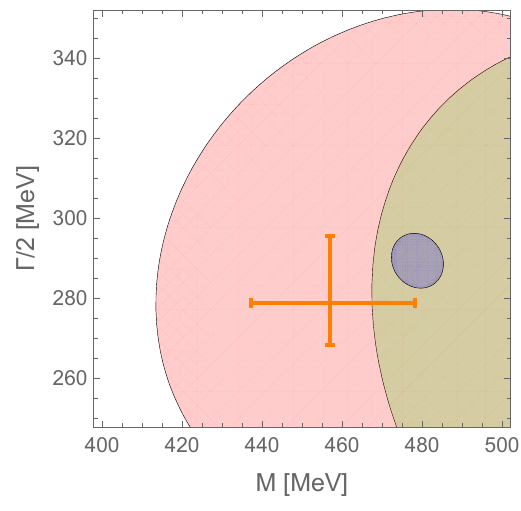}\\
    \small (d) $v_{4}$
  \end{minipage}\hfill
  \begin{minipage}[b]{0.19\textwidth}
    \centering
    \includegraphics[width=\linewidth]{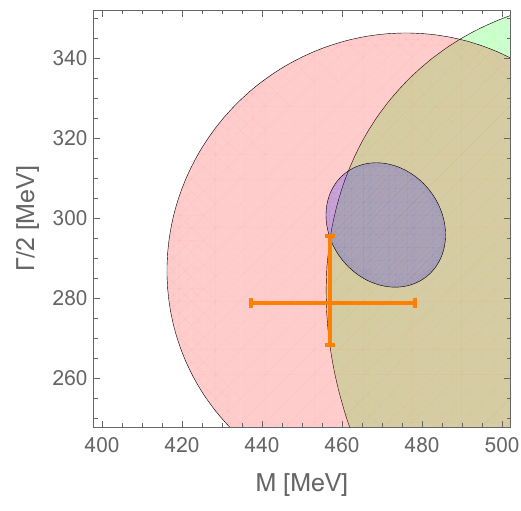}\\
    \small (e) $v_{5}$
  \end{minipage}

  \caption{Overlap of the 68\% CL ellipses for the $P_{2}^{N}$'s sequence pole position: red for $P_{2}^{1}$, green for $P_{2}^{2}$, and blue for $P_{2}^{3}$. Panels a) to e) correspond to parameterizations $v_1-v_5$. The orange cross is the reference number $s_p=(457^{+14}_{-13} - i 279^{+11}_{-7})$.}
  \label{P2iParametrization}
\end{figure*}
 
\section{Pole determinations from Padé Approximants}
\noindent The accuracy of our pole position determinations relies on precise knowledge of the Taylor expansion of $F(s)$ around $s_0$. In Ref.~\cite{Caprini:2016uxy}, five distinct parameterizations, $v_1$--$v_5$, are provided for the scalar isoscalar $\pi\pi$ phase shift $\delta_0^0(s)$, along with its first four derivatives. These parameterizations allow us to compute both the values and derivatives of the $IJ=00$ $\pi\pi$ partial wave given by
\begin{equation}
t_j^I(s)=\frac{e^{2i\delta_J^I(s)}-1}{2i\rho(s)}
\end{equation}
where $\rho(s)=\sqrt{1-4m_{\pi}^2/s }$ is the phase-space factor. In this letter, we extend the analysis carried out in Ref. \cite{Caprini:2016uxy} by explicitly enforcing the physical condition that the imaginary part of the amplitude vanishes at the pion-pion threshold in the construction of the PAs. Applying this condition allows the use of one-order higher PAs without modifying the input data or parameters of Ref. \cite{Caprini:2016uxy}.

For the set of parameterizations, we report the central values of the input parameters and the associated uncertainties, which originate both from the input data and from the truncation of the PA sequence. The pole positions are expressed in terms of the resonance mass and width,
\begin{equation}
    \sqrt{s_p} = M - i \frac{\Gamma}{2}.
\end{equation}

Using the central values and higher-order derivatives for each parameterization provided in Table II of Ref.~\cite{Caprini:2016uxy}, we carried out a systematic study of the convergence properties of the PA sequences: $P_{1}^{N}$ (up to $P_{1}^{4}$) and $P_{2}^{N}$ (up to $P_{2}^{3}$), both sequences chosen to allow a direct comparison with the previous analysis \cite{Caprini:2016uxy,Masjuan:2013jha,Masjuan:2014psa}. The study covers a range of expansion points $s_0$ within the elastic region, from the $\pi\pi$ threshold to 0.85~GeV. The convergence of the sequences is illustrated in Figs.~\ref{P1iParametrization} and \ref{P2iParametrization}, showing that, within each sequence, the most stable determinations are obtained from the highest-order approximants, namely $P_{1}^{4}(s,s_0)$ and $P_{2}^{3}(s,s_0)$, respectively.

As in Ref.~\cite{Caprini:2016uxy}, the truncation uncertainty is defined from the difference between consecutive PAs, and the optimal expansion point $s_0^{\text{opt}}$ is determined by minimizing the combined theoretical and statistical uncertainties through a Monte Carlo procedure. Since the methodology is identical to Refs. \cite{Caprini:2016uxy,Masjuan:2013jha,Masjuan:2014psa}, the technical details are omitted here.

For each parametrization, we then constructed the PAs $P_{1}^{4}(s, s_0^{\text{opt}})$ and $P_{2}^{3}(s, s_0^{\text{opt}})$ and extracted the corresponding pole positions using the central input values. These results, displayed in Tables~\ref{P41} and \ref{P23}, are organized as follows: the second column contains the central determinations of the pole positions, while the third and fourth columns list the theoretical uncertainties from the truncation of the PA sequences and the statistical uncertainties obtained from the Monte Carlo analysis. The final uncertainty, shown in the fifth column, results from the quadratic combination of both sources. In all cases, the mean values of the Monte Carlo distributions differ from the central determinations by less than $1~\text{MeV}$, ensuring the internal consistency of the procedure and confirming the adequacy of the Gaussian approximation for statistical fluctuations.

\begin{figure}[h!]
\begin{adjustwidth}{1 cm}{}
    \subfigure[ \, 68$\%$ CL regions for the $P_{1}^{4}$ pole positions.]{ \includegraphics[width=0.35\textwidth]{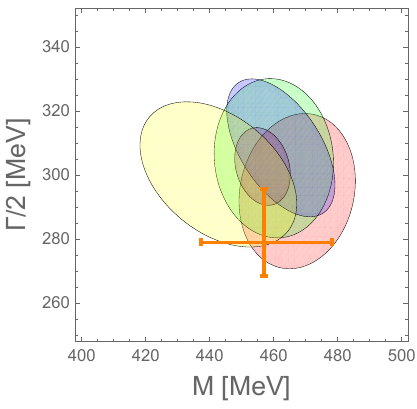}\label{densityregion1}}
    \subfigure[ \,
    68$\%$ CL regions for the $P_{2}^{3}$ pole positions.
        ]{    \includegraphics[width=0.35\textwidth]{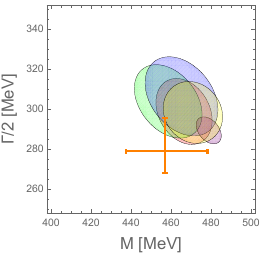}\label{densityregion2}}
\caption{\footnotesize Illustration of the analytic continuation's sensitivity of the PA method by showing the 68\% confidence level (CL) regions for the pole positions obtained from the different parameterizations. The colors correspond to the five parameterizations as follows: red for $v_{1}$, green for $v_{2}$, blue for $v_{3}$, purple for $v_{4}$, and yellow for $v_{5}$. The orange cross is the reference number $s_p=(457^{+14}_{-13} - i 279^{+11}_{-7})$.}
\label{P2iParametrizations}
\end{adjustwidth}
\end{figure}

Figs. \ref{densityregion1} and \ref{densityregion2} further illustrate the sensitivity of the analytic continuation by displaying the 68$\%$ confidence level (CL) regions for the pole positions obtained from the different parameterizations. The overlap of these regions indicates that, although the parameterizations are nearly indistinguishable in the physical region, as they all come from a fit to the same data, and equally satisfy the dispersive constraints, their analytical form yields slightly different pole locations. This behavior highlights the method’s sensitivity to the choice of input and underscores the potential to underestimate theoretical uncertainties when relying on a single parameterization.

To mitigate this issue, we adopt a more realistic uncertainty estimation based on the spread of pole positions obtained from physically equivalent inputs, following the philosophy of previous works \cite{Caprini:2008fc,  Yndurain:2007qm}. In this approach, uncertainties are quantified by examining the variation among results derived from input that is indistinguishable within the physical region. By taking the extremes of the one-standard-deviation ranges reported in Tables \ref{P41} and \ref{P23}, we obtain conservative uncertainty intervals for the $P_{1}^{4}$ and $P_{2}^{3}$, respectively:

\begin{align}
M &= (459.7 \pm 28.9)\textup{ MeV}, \notag \\
\Gamma/2 &= (300.3 \pm 21.3)\textup{ MeV}. \label{firstres4}
\end{align}
\begin{align}
M &= (468.0 \pm 17.8)\textup{ MeV}, \notag \\
\Gamma/2 &= (299.5 \pm 17.2)\textup{ MeV}, \label{firstres1}
\end{align}

It is noteworthy that the $M=2$ approximants produce more stable and accurate results than the $M=1$ case. This improvement arises because, for $M=2$, the additional pole allows for a separation of the resonance dynamics from other analytic structures. In contrast, an $M=1$ approximant must attempt to reproduce both the background effects and the resonance with a single pole term, which slows convergence and increases sensitivity to small variations. By employing $M=2$, one pole accurately locates the resonance, while the other pole effectively parametrizes the background and the influence of the nearby branch cut present in the input data mimicking mathematically all these structures. This happens independently of whether the parameterization would contain or not a second branch cut if the data used to build them do have such information. This results in more stable and precise determinations of the $f_0(500)$-pole parameters.

\begin{table*}[]
\centering
\caption{ Mass and width (in MeV) for the $P_{1}^{4}$(s) approximant and its uncertainty.}
\label{P41}
\begin{tabular}{c|cccc}
\hline
                     & pole               & theo. uncert. & stat. uncert. & combined uncert. \\ \hline
$v_{1}$              & $M=467.4$          & $11.5$      & $3.6$       & $12.0$         \\
($\sqrt{s_{0}}=610$) & $\Gamma/2=295.0 $ & $11.5$      & $11.1$      & $16.0$         \\ \hline
$v_{2}$              & $M=460.0$          & $12.0$      & $3.0$       & $12.3$         \\
($\sqrt{s_{0}}=610$) & $\Gamma/2=305.2$  & $12.0$      & $11.2$      & $16.4$         \\ \hline
$v_{3}$              & $M=471.8$          & $11.9$      & $3.0$       & $12.3$         \\
($\sqrt{s_{0}}=590$) & $\Gamma/2=298.3$  & $11.9$      & $4.7$       & $12.8$         \\ \hline
$v_{4}$              & $M=456.5$          & $3.9$       & $4.2$       & $5.8$          \\
($\sqrt{s_{0}}=580$) & $\Gamma/2=302.7$  & $3.9$       & $7.0$       & $8.0$          \\ \hline
$v_{5}$              & $M=442.7$          & $11.7$      & $11.4$      & $16.3$         \\
($\sqrt{s_{0}}=610$) & $\Gamma/2=300.2$  & $11.7$      & $9.3$       & $15.0$         \\ \hline
\end{tabular}
\end{table*}

\begin{table*}[]
\centering
\caption{ Mass and width (in MeV) for the $P_{2}^{3}$(s) approximant and its uncertainty.}
\label{P23}
\begin{tabular}{c|ccccl}
\hline
                     & pole               & theo. uncert. & stat. uncert. & combined uncert. & 2nd pole                    \\ \hline
$v_{1}$              & $M=466.5$          & $7.0$       & $6.4$       & $9.5$          & $M_{2}=440.7\pm 21.4$        \\
($\sqrt{s_{0}}=430$) & $\Gamma/2=295.0 $ & $7.0$       & $8.2$       & $10.8$         & $\Gamma_{2}/2=11.6\pm 21.4$ \\ \hline
$v_{2}$              & $M=458.6$          & $10.0$      & $5.0$       & $11.2$         & $M_{2}=428.3\pm 24.4$        \\
($\sqrt{s_{0}}=420$) & $\Gamma/2=304.0$  & $10.0$      & $6.8$       & $12.1$         & $\Gamma_{2}/2=13.4\pm 24.4$ \\ \hline
$v_{3}$              & $M=465.2$          & $11.2$      & $4.4$       & $12.1$         & $M_{2}=430.8\pm 20.8$        \\
($\sqrt{s_{0}}=430$) & $\Gamma/2=306.8$  & $11.2$      & $6.5$       & $13.0$         & $\Gamma_{2}/2=6.7\pm 20.8$ \\ \hline
$v_{4}$              & $M=472.4$          & $1.5$       & $3.9$       & $4.2$          & $M_{2}=443.3\pm 7.9$        \\
($\sqrt{s_{0}}=440$) & $\Gamma/2=289.5$  & $1.5$       & $4.2$       & $4.5$          & $\Gamma_{2}/2=-0.3\pm 7.9$ \\ \hline
$v_{5}$              & $M=470.9$          & $7.2$       & $6.5$       & $9.7$          & $M_{2}=464.5\pm 13.9$        \\
($\sqrt{s_{0}}=460$) & $\Gamma/2=298.4$  & $7.2$       & $7.3$       & $10.3$         & $\Gamma_{2}/2=-1.69\pm 13.9$ \\ \hline
\end{tabular}
\end{table*}

Furthermore, in our analysis of the PAs with two poles ($M=2$), the second one is found almost canceled by a zero in its numerator, thus forming a Froissart doublet \cite{BakerGraves-Morris1996}. The accumulation of such doublets is the mechanism by which rational approximants mimic a branch cut. Such behavior is then the expected one when PAs, which are ratios of polynomials, must approximate a function with a branch cut. The lack of higher-order derivatives prevents the analysis of $M>2$ sequences.

A third source of uncertainty can be incorporated, arising from the spread of the central values across different parameterizations, in addition to the theoretical and statistical uncertainties. By taking the average of these predictions as the final central value and combining all three uncertainties in quadrature, we obtain slightly smaller overall uncertainties than those reported in Eqs.(\ref{firstres4}) and (\ref{firstres1}),  
\begin{align}
M &= (459.7 \pm 16.0)\,\text{MeV}, \notag \\
\Gamma/2 &= (300.3 \pm 13.9)\text{MeV}, \label{secondres1}
\end{align}
\begin{align}
M &= (468 \pm 12.4)\,\text{MeV}, \notag\\
\Gamma/2 &= (299.5 \pm 11.9) \text{MeV}. \label{secondres4}
\end{align}

Compared to previous analyses \cite{Caprini:2016uxy,Masjuan:2013jha,Masjuan:2014psa}, the uncertainties in the extracted pole parameters are now substantially reduced. For the $P_1^4$, the uncertainty of the mass decreases by approximately 27\%, and the uncertainty of the half-width by roughly 44\%. For the $P_2^3$, the mass uncertainty is reduced by approximately 41\%, while the half-width shows a decrease of approximately 21\%. This more robust behavior is expected for PAs subject to threshold constraints: by incorporating physical information at the pion threshold, the analytic structure is improved and the sequence converges more rapidly. Moreover, this result aligns with theoretical expectations, as imposing additional constraints typically necessitates using higher-order $N$ in the PA sequence, which naturally reduces the associated theoretical uncertainty and returns smaller final uncertainties. We further illustrate the comparison in the last section of this letter.

\section{Pole Determinations from Padé approximants Using Updated Parameterizations}
\begin{figure}[h!]
\begin{adjustwidth}{0.2cm}{}
    \subfigure[\, 68\% CL ellipses for the $v_6$ $P_{1}^{N}$'s sequence pole position.]{ \includegraphics[width=0.35\textwidth]{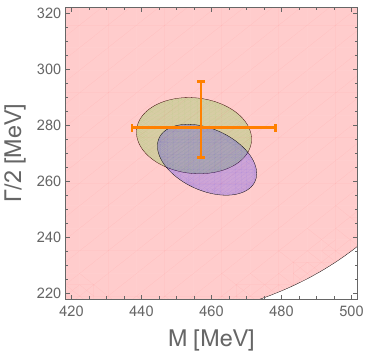}\label{v6p1n}}
    \subfigure[\, 68\% CL ellipses for the $v_6$ $P_{2}^{N}$'s sequence pole position.
        ]{    \includegraphics[width=0.35\textwidth]{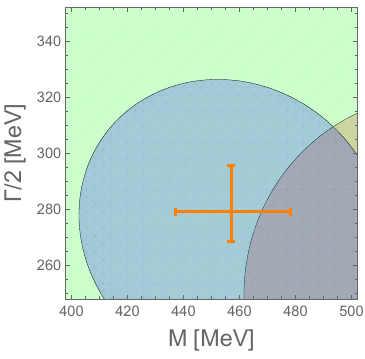}\label{v6p2n}}
\caption{\footnotesize Overlap of the 68\% CL ellipses for the pole positions derived from the $P_1^N$ ($P_1^2$–$P_1^4$), upper panel, and $P_2^N$ ($P_2^1$–$P_2^3$), lower panel. Red, green, and blue colors correspond to increasing PA order, respectively. The orange cross is the reference number $s_p=(457^{+14}_{-13} - i 279^{+11}_{-7})$.}
\label{P2iParametrizationsv6}
\end{adjustwidth}
\end{figure}

\begin{figure}[]
\begin{adjustwidth}{1 cm}{}
    \subfigure[\, 68$\%$ CL regions for the pole positions from $P_{1}^{4}$ for all parameterizations.]{ \includegraphics[width=0.35\textwidth]{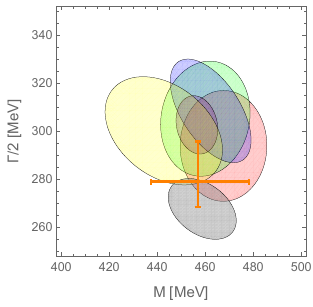}\label{densityregion1v6}}
    \subfigure[\, 68$\%$ CL regions for the pole positions from $P_{2}^{3}$ for all parameterizations.
        ]{    \includegraphics[width=0.35\textwidth]{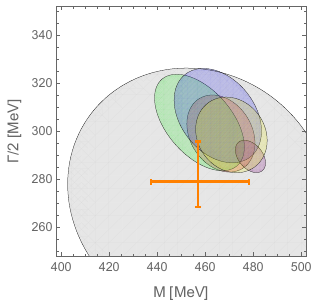}\label{densityregion2v6}}
\caption{\footnotesize Overlap of the 68\% CL ellipses for $P^4_1$ (upper panel) and $P^3_2$ (lower panel). The colors correspond to the six parameterizations as follows: red for $v_{1}$, green for $v_{2}$, blue for $v_{3}$, purple for $v_{4}$, yellow for $v_{5}$ and gray for $v_{6}$.  The orange cross represents the reference value $s_p = (457^{+14}_{-13} - i 279^{+11}_{-7})$.}
\label{P2iParametrizationsall}
\end{adjustwidth}
\end{figure}

 \begin{figure*}[]
    \centering
    \includegraphics[width=0.8\textwidth]{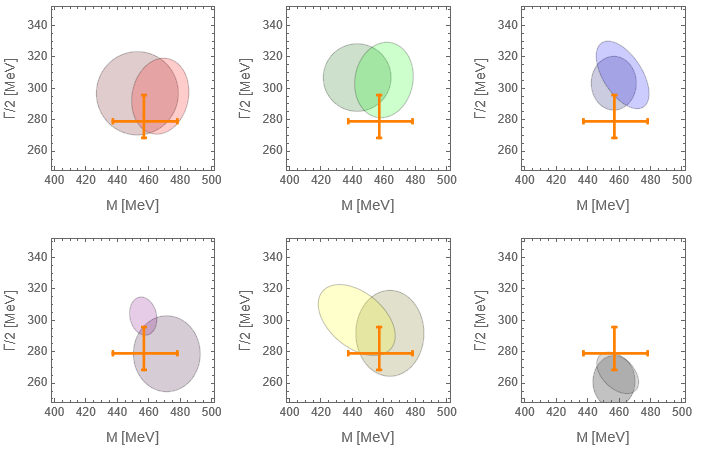} 
\caption{Comparison of the 68\% confidence level (CL) ellipses for the six parameterizations. The darker ellipses correspond to the results obtained from Ref. \cite{Caprini:2016uxy} for $P_1^3$, while the lighter ellipses show the updated results presented in this work for $P_1^4$. Colors indicate the different parameter sets: red for $v_1$, green for $v_2$, blue for$v_3$, purple for $v_4$, yellow for $v_5$, and gray for $v_6$. The orange cross is the reference number $s_p=(457^{+14}_{-13} - i 279^{+11}_{-7})$.}
    \label{CompPereandmeP14}
\end{figure*}

\begin{figure*}[]
    \centering
    \includegraphics[width=0.8\textwidth]{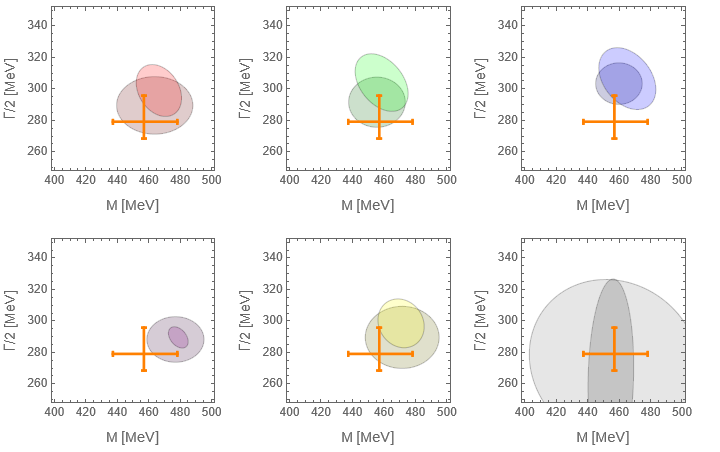} 
\caption{Comparison of the 68\% confidence level (CL) ellipses for the six parameterizations. The darker ellipses correspond to the results obtained from Ref.\cite{Caprini:2016uxy} for $P_2^2$, while the lighter ellipses show the updated results presented in this work for $P_2^3$. Colors indicate the different parameter sets: red for $v_1$, green for $v_2$, blue for $v_3$, purple for $v_4$, yellow for $v_5$, and gray for $v_6$. The orange cross is the reference number $s_p=(457^{+14}_{-13} - i 279^{+11}_{-7})$.}
    \label{CompPereandmeP23}
\end{figure*}

\noindent In this section, we employ the updated version of the $v_1$ parametrization introduced in Ref.~\cite{Pelaez:2019eqa}. For clarity, we denote this new parameterization by $v_6$, and use the set of free parameters (together with their uncertainties) provided in  Table XI of Appendix B of Ref.~\cite{Pelaez:2024uav}. Following the strategy outlined above, we examine the convergence of the $P_1^N$ and $P_2^N$ sequences for this new parametrization.

As shown in Table \ref{v6}, the $v_6$ values are consistent with the updated results reported in Ref.~\cite{Pelaez:2024uav}, which follow from the parametrization derived by the same authors in earlier works.
 
 As illustrated in Figs.~\ref{v6p1n} and \ref{v6p2n}, the sequences converge, though more slowly than for the other parameterizations. This could arise from the fact that, while $v_6$ is structurally analogous to $v_1$, it is contracted using more parameters than before ($N=5$ instead of $N=3$) with a different fit to data, with a slower convergence than before.

\begin{table}[h!]
\caption{Mass and width (in MeV) for the $v_6$ parametrization}
\label{v6}
\hspace{-0.55 cm}
\begin{tabular}{c|cccc}
\hline
                & pole             & th. uncert. & stat. uncert. & combined \\ \hline
$P_1^4(s)$      & $M=458.7$        & 6.6         & 6.7         & 9.4            \\
($\sqrt{s_0}=700$) & $\Gamma/2=260.6$ & 6.6         & 5.1         & 8.4            \\ \hline
$P_2^3(s)$     & $M=457.2$        & 35.0         &  8.3        & 36.0           \\
($\sqrt{s_0}=540$) & $\Gamma/2=275.2$ & 35.0        & 3.6         & 35.2           \\ \hline
\end{tabular}
\end{table}

Figs. \ref{densityregion1v6} and \ref{densityregion2v6} further illustrate the sensitivity of the analytic continuation by displaying the 68\% CL regions for the pole positions obtained from the different parameterizations. We notice that the uncertainty ellipse associated with the $v_6$ parameterization is significantly larger than those of the other parameterizations for the $P_2^N$ sequence, reinforcing the conclusion that its convergence is slower. In contrast, for the $P_1^N$ sequence, the ellipse overlaps with some, but not all, of the other parameterizations.

\section{Role of correct threshold behavior and final determination}

To assess the impact of enforcing the correct threshold behavior, we compare in detail the pole positions obtained in this work with previous results \cite{Caprini:2016uxy}, the main difference being the inclusion of physical constraints at the pion threshold. Figs. \ref{CompPereandmeP14} and \ref{CompPereandmeP23} show the pole determinations for each parameterization $v_1$–$v_6$, with darker ellipses representing the reference results without the threshold \cite{Caprini:2016uxy} and lighter ellipses the updated new results. Including the threshold information allows us to explore one PA higher on each sequence, still using the same amount of information from the scattering amplitude. Globally, it is clear that enforcing the correct threshold returns a better determination of the pole position, even though the inputs from the scattering amplitude are the same and with the same uncertainties. This is especially so for the $P^N_2$ sequence with the exception of $v_6$.

By combining the updated results for all parameterizations, $v_1$–$v_6$, and applying the same uncertainty evaluation procedure as in Eqs. (\ref{secondres1}) and (\ref{secondres4}) of the previous section, we get

\begin{align}
M &= (459.5 \pm 14.9),\text{MeV}, \notag \\
\Gamma/2 &= (294.8 \pm 18.6),\text{MeV}, \label{thirdres2}
\end{align}
for $P_1^4$, and 
\begin{align}
M &= (466.2 \pm 16.0),\text{MeV}, \notag \\
\Gamma/2 &= (295.0 \pm 18.2),\text{MeV}, \label{thirdres1}
\end{align}
for $P_2^3$. Although the $v_6$ parametrization carries larger individual uncertainties, its inclusion still improves the overall determination of the pole parameters \cite{Caprini:2016uxy,Masjuan:2013jha,Masjuan:2014psa}. For $P_2^3$, the mass uncertainty decreases by approximately 24\%, while the half-width increases slightly by 14\%. In the case of $P_1^4$, both the mass and half-width uncertainties are reduced by roughly 32\% and 26\%, respectively. These results demonstrate that imposing the correct threshold behavior in the $\pi\pi$ partial-wave amplitude leads to an overall reduction of the final uncertainties. The threshold constraint not only improves the analytic continuation but also ensures the reliability of the extracted resonance parameters,  rendering the PA method one of the simplest, most efficient, and accurate methods to study resonance properties.

\begin{acknowledgments}
We are deeply grateful to Juan José Sanz-Cillero for his encouragement and support throughout the preparation of this work, as well as for his thoughtful review of the manuscript.

Our work has been supported by the Ministerio de Ciencia e Innovación under grant PID2020-112965GB-I00, by the Secretaria d’Universitats i Recerca del Departament d’Empresa i Coneixement de la Generalitat de Catalunya under grant 2021 SGR 00649, and by the Spanish Ministry of Science and Innovation (MICINN) through the State Research Agency under the Severo Ochoa Centres of Excellence Programme 2025–2029 (CEX2024-001442-S). IFAE is partially funded by the CERCA program of the Generalitat de Catalunya. B.D. also acknowledges support from the predoctoral program AGAUR-FI (2025 FI-3 00065) Joan Oró of the Department of Research and Universities of the Generalitat de Catalunya, co-financed by the European Social Fund Plus.

\end{acknowledgments}

\nocite{*} 

\end{document}